\documentclass[natbib]{svjour3}                     
\smartqed  
\usepackage{graphicx, psfig}
\usepackage{amssymb}

 \usepackage{aps-bibstyle}  
\newcommand{\fesc}{\mbox{$f_{\rm esc}$}}

\def\loz{$\ell(z)$}

\def\CIVdblt{{\rm C~}\kern 0.1em{\sc iv}~$\lambda\lambda 1548, 1550$}
\def\MgIIdblt{{\rm Mg~}\kern 0.1em{\sc ii}~$\lambda\lambda 2796, 2803$}
\def\NVdblt{{\rm N}\kern 0.1em{\sc v}~$\lambda\lambda 1238, 1242$}  
\def\OVIdblt{{\rm O}\kern 0.1em{\sc vi}~$\lambda\lambda 1031, 1037$}
\def\SiIVdblt{{\rm Si~}\kern 0.1em{\sc iv}~$\lambda\lambda1394, 1403$}
\def\AlIIIdblt{{\rm Al~}\kern 0.1em{\sc iii}~$\lambda\lambda1855,1863$}
\def\FeIIdblt{{\rm Fe~}\kern 0.1em{\sc ii}~$\lambda\lambda 2383, 2600$}

\def\AlII{\hbox{{\rm Al~}\kern 0.1em{\sc ii}}}
\def\AlIII{\hbox{{\rm Al~}\kern 0.1em{\sc iii}}}
\def\CaI{\hbox{{\rm Ca}\kern 0.1em{\sc i}}}
\def\CaII{\hbox{{\rm Ca}\kern 0.1em{\sc ii}}}

\def\CrII{\hbox{{\rm Cr~}\kern 0.1em{\sc ii}}}
\def\CI{\hbox{{\rm C~}\kern 0.1em{\sc i}}}
\def\CII{\hbox{{\rm C~}\kern 0.1em{\sc ii}}}
\def\CIII{\hbox{{\rm C~}\kern 0.1em{\sc iii}}}
\def\CIV{\hbox{{\rm C~}\kern 0.1em{\sc iv}}}
\def\CV{\hbox{{\rm C}\kern 0.1em{\sc v}}}

\def\H{\hbox{{\rm H}}}
\def\HI{\hbox{{\rm HI}}}
\def\H2{\hbox{{\rm H$_2$}}}

\def\HII{\hbox{{\rm H}\kern 0.1em{\sc ii}}}

\def\Lya{\hbox{{\rm Ly}\kern 0.1em$\alpha$}}

\def\Lyb{\hbox{{\rm Ly}\kern 0.1em$\beta$}}
\def\Lyg{\hbox{{\rm Ly}\kern 0.1em$\gamma$}}
\def\Lyfive{\hbox{{\rm Ly}\kern 0.1em$5$}}
\def\Lysix{\hbox{{\rm Ly}\kern 0.1em$6$}}
\def\Lyseven{\hbox{{\rm Ly}\kern 0.1em$7$}}
\def\Lyeight{\hbox{{\rm Ly}\kern 0.1em$8$}}
\def\Lynine{\hbox{{\rm Ly}\kern 0.1em$9$}}
\def\Lyten{\hbox{{\rm Ly}\kern 0.1em$10$}}
\def\HeI{\hbox{{\rm He}\kern 0.1em{\sc i}}}
\def\HeII{\hbox{{\rm He}\kern 0.1em{\sc ii}}}
\def\FeI{\hbox{{\rm Fe~}\kern 0.1em{\sc i}}}
\def\FeII{\hbox{{\rm Fe~}\kern 0.1em{\sc ii}}}
\def\FeIII{\hbox{{\rm Fe~}\kern 0.1em{\sc iii}}}
\def\MnII{\hbox{{\rm Mn}\kern 0.1em{\sc ii}}}
\def\MgI{\hbox{{\rm Mg~}\kern 0.1em{\sc i}}}
\def\MgII{\hbox{{\rm Mg~}\kern 0.1em{\sc ii}}}
\def\MgIII{\hbox{{\rm Mg~}\kern 0.1em{\sc iii}}}
\def\MgIV{\hbox{{\rm Mg~}\kern 0.1em{\sc iv}}}
\def\NaI{\hbox{{\rm Na}\kern 0.1em{\sc i}}}
\def\NV{\hbox{{\rm N}\kern 0.1em{\sc v}}}
\def\NII{\hbox{{\rm N}\kern 0.1em{\sc ii}}}
\def\NIII{\hbox{{\rm N}\kern 0.1em{\sc iii}}}
\def\NiI{\hbox{{\rm Ni~}\kern 0.1em{\sc i}}}
\def\NiII{\hbox{{\rm Ni~}\kern 0.1em{\sc ii}}}
\def\OVI{\hbox{{\rm O}\kern 0.1em{\sc vi}}}
\def\OI{\hbox{{\rm O}\kern 0.1em{\sc i}}}
\def\OII{\hbox{[{\rm O}\kern 0.1em{\sc ii}]}}
\def\OIII{\hbox{[{\rm O}\kern 0.1em{\sc iii}]}}
\def\SiII{\hbox{{\rm Si~}\kern 0.1em{\sc ii}}}
\def\SiIII{\hbox{{\rm Si~}\kern 0.1em{\sc iii}}}
\def\SiIV{\hbox{{\rm Si~}\kern 0.1em{\sc iv}}}
\def\SII{\hbox{{\rm S~}\kern 0.1em{\sc ii}}}
\def\SIII{\hbox{{\rm S~}\kern 0.1em{\sc iii}}}
\def\SIV{\hbox{{\rm S~}\kern 0.1em{\sc iv}}}
\def\TiII{\hbox{{\rm Ti}\kern 0.1em{\sc ii}}}
\def\ZnII{\hbox{{\rm Zn~}\kern 0.1em{\sc ii}}}

\def\xray{X-ray}
\def\swift{\emph{Swift}}

\def\simlt{\mathrel{\hbox{\rlap{\hbox{\lower4pt\hbox{$\sim$}}}\hbox{$<$}}}}
\def\simgt{\mathrel{\hbox{\rlap{\hbox{\lower4pt\hbox{$\sim$}}}\hbox{$>$}}}}

\newcommand{\nh}{\mbox{$N_{\rm HI}$}} 

\begin{document}


\title{GRBs as Probes of the IGM 
}


\author{Antonino Cucchiara         \and
        Tonomori Totani\and
        Nial Tanvir 
}


\institute{Antonino Cucchiara \at
              Space Telescope Science Institute,\\ 
              3700 San Martin Drive, Baltimore \\
              Tel.: +1-410-338-6759\\
              \email{acucchiara@stsci.edu}           
           \and
           Tonomori Totani \at
           Department of Astronomy, \\
           The University of Tokyo, \\
           Hongo, Tokyo 113-0033   
           \and
           Nial Tanvir \at
           Department of Physics and Astronomy,\\ 
           University of Leicester, University Road,\\
           Leicester, LE1 7RH, UK
}

\date{Received: date / Accepted: date}

\maketitle

\begin{abstract}
Gamma-ray Bursts (GRBs) are the most powerful explosions known, capable of outshining the rest of gamma-ray sky during their short-lived prompt emission. Their
cosmological nature makes them the best tool to explore the final stages in the lives of very massive stars up to the
highest redshifts. 
Furthermore, studying the emission
from their low-energy counterparts (optical and infrared) via rapid spectroscopy, we have been able to pin down the exact location of the most distant galaxies as well as placing stringent constraints on 
 their host galaxies and intervening systems  at low and high-redshift (e.g. metallicity and neutral hydrogen fraction).
In fact, each GRB spectrum contains absorption features imprinted by metals in the host
interstellar medium (ISM)  as well as the intervening intergalactic medium (IGM)  along the line of sight. 
In this chapter we summarize the progress made using a large dataset of GRB spectra in understanding the nature of both these absorbers and how
GRBs can be used to study the early Universe, in particular to measure the neutral hydrogen fraction and the escape fraction of UV photons before and during the epoch of re-ionization.

\end{abstract}

\section{Introduction}
\label{intro}
Gamma-ray Bursts (GRB) have been observed from $z=0.03$ \citep{Pian:2006aa}  to $z\sim9$ \citep{Cucchiara:2011lr,sdc09,tfl09} and, similar to quasars (QSOs), their X-ray/optical/NIR afterglows
intersect the interstellar medium (ISM) and intergalactic medium (IGM) along their lines of sight \citep[e.g. ][]{Jakobsson:2004aa,Behar:2011aa,Jakobsson:2012aa,Starling:2013aa,Campana:2015aa}: 
from matter near to the host  \citep[local ISM,][]{Zafar:2010aa} to other galaxies (intervening ISM and/or IGM) at lower redshift (and different impact parameters). The gas located in each of these components leaves an imprint clearly identified in GRB (and QSO) optical and NIR spectra by the presence of absorption lines.

Therefore, absorption spectroscopy allows us to disentangle all these absorption 
systems allowing us to discern not only their properties 
(e.g. metal content and gas kinematics of these galaxies), but also the overall characteristics of the cosmic IGM during and after the end of re-ionization 
in a very similar fashion to QSOs \citep[see][for a review]{Fan:2006aa}.

In the following sections we summarize the advantage of using GRB afterglows
 for these studies with respect to QSOs, the current status of studying the chemical evolution of the Universe using GRB intervening systems, and  the investigation of the cosmic IGM (e.g. neutral hydrogen fraction)  based on recent 
 GRB spectroscopic datasets. We will also emphasize some of the drawback of using GRBs, 
 e.g., their transient nature and the 
 low number of events in comparison with QSOs over most redshifts.

\section{Intervening systems}
\label{sec:1}
Quasar lines of sight have been explored in depth in recent years thanks to the availability of 
large spectroscopic surveys, like the Sloan Digital Sky Surveys \citep{York:2000uq}.
Since they are bright background sources, they probe gas and matter located in foreground objects at different impact parameters. These intervening systems can be identified in the QSO spectra based on metal absorption features at redshifts lower than the QSO systemic redshift (usually determined by  \Lya, \CIV, and \MgII\ emission lines). However, the complexity of the intrinsic QSO continuum and the decreased numbers of bright QSOs at high-redshift makes the study of the intervening systems challenging.
Nevertheless, thanks to several observing campaigns spanning the near-UV to the near-Infrared, a large sample of such objects has been collected: the identification of \Lya\ lines at the intervening system redshifts has offered a unique opportunity to use them as tracers of the metal enrichment of the Universe from low-redshift \citep{Lehner:2013aa,Muzahid:2016aa,Cooper:2015aa} to $z\sim5$. Furthermore, \MgII\ intervening absorbers have been identified in several QSO spectra: in the last decade the presence and metal content of such absorbers have been matter of debate, in particular regarding the role of such absorbers in fuelling star-formation and/or tracing cosmological galaxy build up. 

Both types of absorbers, \Lya\ and \MgII, have been also identified in GRB afterglow spectra, the former
indicating the presence of large reservoir of neutral hydrogen in the GRB host galaxy, while the latter 
are often produced by foreground absorbers along the GRB line of sight. 
The synchrotron emission responsible for the simple power-law afterglow continuum makes the identification and measurement of absorption lines easier than QSOs, but the rarity and rapidly fading nature of the
emission means that much lower numbers of such systems have been identified. 
Nevertheless, GRBs have been spectroscopically identified up to $z=8.2$, 
providing unprecedented tool to study the epoch of re-ionization.
Thanks largely to the sample of GRBs discovered by the \swift\ satellite \citep{ger04}, and the  identification of their afterglows
in many cases (from \xray\ to optical and NIR), we have been able to 
investigate samples of  these absorbers in considerable detail.
In the following sub-sections we will briefly discuss these two types of absorbers, though we direct the reader to the \emph{Host galaxy} chapter for a more complete view of the GRB hosts properties.

\subsection{Lyman-$\alpha$ absorbers}
Intervening systems for which a \Lya\ line is identified in absorption are subdivided based on the total 
hydrogen column density: if ${\rm log\;} \nh \geq 20.3 \;{\rm cm}^{-2}$ they are called ``Damped Lyman-$\alpha$ systems'' (DLAs), while if ${\rm log\;} \nh < 20.3 \;{\rm cm}^{-2}$, the systems are generically called sub-DLAs  \citep[see ][for a detailed review on such absorbers and further subdivisions]{wgp05}.
This cool gas usually surrounds galaxies and can be used to characterize the circumgalactic (CGM) medium   
\citep{Stewart:2011aa,Stinson:2012aa,Rudie:2012aa} testing theoretical predictions 
\citep[e.g. ][]{Nagamine:2004aa,Berry:2014aa,Cen:2012aa,Razoumov:2008aa,Fumagalli:2011aa}.
Because of their connection with star-formation, DLAs represent important laboratories to chart the metal
enrichment of the Universe, from the end of re-ionization to the peak of star-formation ($z\sim2$) and at 
lower redshifts. Recently, \citet{Rafelski:2014aa} have shown how metallicity decreases with redshift up to  $z\sim4.5$ in a large QSO-DLA sample. Furthermore, these authors suggest that  at even higher redshifts there is a steepening in the decline of cosmic metallicity, probably due to an increase of the covering fraction of neutral gas as function of redshift.
\Lya\ absorbers have also been identified in the GRB afterglow spectra and offer a unique possibility to obtain not only higher signal-to-noise ratio spectra (bright GRB afterglows observed early are usually brighter than QSOs at any redshift) but also to trace the stellar content of such DLAs (GRB-DLAs): in fact, GRB-DLAs are associated with the host galaxies enabling, once the afterglow fades away, detailed studies of DLAs itself.

In 2015, \citet{Cucchiara:2015aa}  compiled the largest sample of spectroscopic afterglow data available and determined uniformly the metallicities of 54 GRB-DLAs from weak metal absorption lines (like \SII\, \SiII). They compared this sample with the QSO-DLA results from Rafelski et al. and, despite having a smaller sample, showed a higher metallicity around GRB-DLAs at $z=3-5$, roughly 10\% the solar value (a factor of two higher than the one measured for QSO-DLAs). These authors suggest that GRB-DLAs may be the best tracer of the interconnection between DLAs and metal build up, especially at high redshift since usually GRBs occur within their host, in close proximity to star-forming regions.
The host galaxy metallicity determination obtained by absorption lines diagnostics is an important aspect 
of cold gas fuelling star-formation and the advent of NIR spectrographs will enable comparisons with 
emission line (integrated over the whole galaxy) diagnostics, a critical step forward on our understanding 
of star-formation in the local Universe and at the highest redshifts.
Furthermore, understanding the nature of the galaxies hosting these absorbers will provide insights on galaxy formation and evolution. The data required to perform such analyses (e.g. extensive multiband observations) is demanding 
because of the faint nature of the sources and the need for space-based observatory time in many cases.

\subsection{\MgII\ absorbers}

One of the most studied and commonly identified feature is the \MgII\ doublet at $\lambda\lambda 2796,2803$\AA, thanks to its large rest wavelength (which makes it them detectable at $z=0.5-2.2$),
the relatively high abundance, and its oscillator strength.

The \MgII\ systems are usually classified in terms of 
the rest-frame equivalent width, $W_r$, of the bluer component 
as ``weak'' ($W_{2796} <0.3$ \AA), 
``strong'' ($W_{2796} >0.3$ \AA) as in \citet{Steidel:1992kx} and \citet{Churchill:1999aa},
and ``very strong''  \citep[$W_{2796} >1.0$ \AA, like in][]{Rodriguez-Hidalgo:2012aa}.


In the last few years several surveys have expanded our samples of \MgII\ intervening systems up to 
$z = 5.2$, thanks to near-IR spectroscopy 
\citep[e.g.][]{Steidel:1992kx,Nestor:2005fk,Prochter:2006kx,
Quider:2011fk,Simcoe:2011uq, Zhu:2013aa}.
These surveys indicate that the very strong \MgII\ absorbers present an increasing
trend up to $z\sim3$ before declining at higher redshift \citep[see Figure 1,][]{Prochter:2006kx,Matejek:2012bh}.
Intriguingly, this behavior closely tracks the cosmic star formation history
\citep{Prochter:2006kx,Zhu:2013aa}, suggesting that some systems may be causally connected to
on-going star formation \citep{Menard:2011vn,Matejek:2012bh}.

For many years, strong \MgII\ absorbers have been associated with galaxies ($L \approx L^*$)
at modest impact parameter and gas around them \citep{Bergeron:1986fk,Lanzetta:1987vn,Steidel:1993fk}, either in the outer disks and/or the circumgalactic medium (CGM).

More recently, QSO lines of sight have been explored in order to probe the 
baryon content around low-$z$ galaxies \citep[][and reference therein]{Kacprzak:2012kx,Chen:2008dq}, 
as well as a diagnostic of the inner part of these galaxies' 
interstellar medium \citep{Bowen:1995bh}.

In a similar fashion, Gamma-ray Bursts (GRBs) provide both information on their hosts
and on the intercepting matter along their lines of sight \citep{Metzger:1997lr}.
There are two main advantages of using these powerful sources: 
first, they can be observed up to very high redshifts  
\citep{Kawai:2006fk,tfl09,sdc09,Cucchiara:2011lr},
which allows one to explore a larger redshift path length, and second, their discovery 
is largely unbiased with respect to intrinsic properties of their hosts 
(extinction, luminosity or mass). In fact, the detection GRBs, at least in \xray\
is independent (or substantially less affected) by the presence of dust extinction along the line
of sight (in the host or in the IGM)  \citep{Zafar:2010aa,Watson:2012aa,Watson:2013aa,Schady:2010aa}

Unfortunately, the number of GRBs discovered and observed  \citep[thanks also to the 
\swift\ satellite,][]{Gehrels:2004fj,Gehrels:2009fk} is several orders of magnitude
less than the number of quasars that have been found in large surveys.
Nevertheless, in 2006
a survey of \MgII\ absorbers
 was performed for an early
sample of {\it Swift} bursts and a heterogeneous sample of pre-existing GRB
spectra \citep[][P06 hereafter]{Prochter:2006fr}.
These authors revealed an extremely puzzling result: the incidence of
strong ($W_{2796} \ge 1$\AA) intervening \MgII\ absorbers was about 4 times
higher along GRB sightlines than quasar sightlines.
No such excess was found in other common class of
absorbers, e.g. \CIV\ features \citep{Tejos:2007aa,ssv07}, or 
for weak \MgII\ absorbers \citep{Tejos:2009aa}.

Several explanations for this discrepancies were proposed by P06,
including: 1) a possible intrinsic origin of these absorbers, which would imply an average
escape velocity from the GRB hosts of $\sim 10-30\%$ the speed of light \citep{Cucchiara:2009mgii,Bergeron:2011lr}; 
2) a significant dust bias along QSO lines of sight
\citep{mnt08,pvs07,Budzynski:2011uq,Sudilovsky:2009aa}; 
3) a geometric effect difference due to the sizes of the emitting regions 
between GRBs and QSOs \citep{sr97,fbs07,pvs07, Lawther:2012fk} ; 
4) a gravitational lensing effect
\citep{Vergani:2009aa,Porciani:2001vn,Rapoport:2011kx,Rapoport:2013aa}. 

Most of these suggestions were later on discarded, but the puzzling result was only solved 
few years ago: the original P06 work,
and even the studies that followed, relied on
a small sample of GRB afterglow spectra.   Even the largest analysis
used only 26 lines of sight (finding 22 absorbers), 
for a total redshift path of $\Delta z =31.55$ \citep{Vergani:2009fj}.
Furthermore, no study had analyzed a completely
independent set of GRB sightlines from the P06 analysis.

In 2013 \cite{Cucchiara:2013aa} collected the largest compilation of GRB spectroscopic data,
exploring a larger redshift path length ($\Delta z=44.9$, Figure\,2).
No excess was identified in an independent sample and further tests, including Montecarlo simulation and bootstrapping analysis, suggested that the earlier work by P06 was biased by a statistical fluke:
in particular, the presence of a small set of
lines of sight with multiple absorbers appears to have driven the
results \citep[as suggested by][]{Kann:2010fk}. The inclusion of the original P06 dataset
does not provide any significant improvement (Figure\,3).

\section{Neutral hydrogen fraction}
\label{sec:3}
The reionization era describes the time when all the intergalactic hydrogen was first ionized by 
the photons produced by early galaxies and stars. Moreover, the same process heated the IGM gas precluding the formation of sub-haloes, which ultimately would form the smaller dwarf galaxies. Therefore, in order to properly describe the evolution of the universe until the present time, it is critical to be able to characterize this important period of its history, for example how long it lasted and how it proceeded (patchy or continuously).

The intergalactic medium (IGM) is known to have been almost completely ionized since a redshift of $z\sim6$ \citep{Becker:2006aa,Fan:2006aa,Becker:2015aa},
while observations of the electron scattering of the cosmic microwave background radiation suggest
the universe transitioned from a cold neutral IGM to a hot ionized one  at an average redshift of $z\sim9$
\citep[][]{Planck-Collaboration:2015aa}, but with the possibility that it was a significantly extended and 
environment dependent process (we will come back to this issue in Section~\ref{sec:4}).

From the detection of the Gunn-Peterson trough \citep[GP,][]{gp65} using high redshift quasars
the optical depth of the universe to ionizing photons rapidly decreases from $z\sim7$ to $z\sim6$, suggesting
that the IGM changed due to re-ionization.
Counting the fraction of spectral pixels that are completely absorbed \citet{McGreer:2011aa}
were able to place a conservative, almost model independent, lower bound on the filling factor of ionised regions. 
Specifically, first results showed that the neutral hydrogen fraction should be smaller than $f_{\rm HI}<0.2$ (1-$\sigma$ confidence) in the $5\leq z\leq 5.5$ redshift range. Later, \citet{McGreer:2015aa}, 
refined this result obtaining  $f_{\rm HI}<0.11$ at $z=5.6$, and $f_{\rm HI}<0.09$ at $z=5.9$.
Another technique, involves the size distribution of 'dark gaps' in the \Lya\ forest \cite[see for example][]{Songaila:2010aa,Gallerani:2008aa,Mesinger:2010aa}, which shows that reionizaton is indeed an
inhomogeneous process, with $f_{\rm HI}<0.1$ at $z\sim5$.

The large number of quasar spectra now available \citep[largely thanks to SDSS,][]{York:2000uq}, offers an opportunity 
to make such measurements, but unfortunately the signal to noise required to perform accurate measurements
is only achieved in a small fraction of cases. 
Moreover, the GP trough gives only weak lower bounds on the neutral gas fraction and the quasar itself may have altered the ionization of the surrounding IGM \citep{Barkana:2004aa}, making obtaining $f_{\rm HI}$ estimates even more complicated.

A new possibility to study the IGM and reionization comes by using high-redshift GRBs: the typical GRB afterglow intrinsic synchrotron emission can be fit by a power-law enabling better constraints on the 
red damping wing of  \Lya\ than is possible for quasars\citep{mr98b,lr00,Ciardi:2000aa,McQuinn:2008aa}. Also GRBs are less biased than QSOs, tracing more typical IGM environments, their radiation has negligible affect on the IGM itself (shorter time-scale radiation than QSOs), and in the first couple of days after their discovery they are much brighter than the average QSO at $z \gtrsim 6$ (e.g. GRB 090423 at $z=8.2$ was $K=20.6$ at 4 days post-burst), with the potential for high-resolution, high signal-to-noise ratio spectroscopy.

For several years, only the spectrum of GRB\,050904\citep[at $z=6.295$][]{Totani:2006aa} was able to provide constraints on the neutral hydrogen fraction
($f_{\rm HI} <0.6 $), based solely on analysis of the red-damping wing. Recently, a newly discovered
GRB at $z=5.912$, GRB\,130606A, has provided a great dataset for a similar analysis \citep{Chornock:2013aa,Totani:2014aa,Hartoog:2015aa}: the bright optical/NIR 
afterglow enabled not only a large follow-up effort, but absorption spectroscopy has shown a low \HI\
column density in the host, implying a much better constraint on the actual $f_{\rm HI}$ of the IGM. 

GRB\,130606A was discovered by the \swift\ and KONUS-\emph{Wind} \citep{Golenetskii:2013aa}
and its redshift was identified based on \HI\ and metal absorption features at
the same redshift ($z=5.912$). The afterglow was observed spectroscopically with several instruments/telescopes:
from the low-resolution FOCAS camera on Subaru telescope (resolving power of $R\sim 900$) to the X-Shooter instrument on VLT
($R\sim10000$). Here we present the results from the former dataset and discuss some discrepancies with other groups' conclusions \citep{Chornock:2013aa,Totani:2014aa,Hartoog:2015aa}.

The FOCAS data were obtained during 9.3-16.5 hours after burst and clearly show the presence of
a strong \Lya\ line. Also, flux calibration was performed using the standard star
Feige 34, obtained the same night.

The observed spectrum was fitted by a model which includes a simple power-law intrinsic 
continuum ($F_{\nu} \propto \nu^{\beta}$) and three components for the \HI\ absorption feature:
\HI\ from the host galaxy (host DLA), a diffuse contribution from the IGM, and a possible intervening Damped \Lya\
system at lower redshift (DLA). 
For the host component a simple Gaussian profile with radial dispersion $\sigma_v$ and
with a column density of $N_{\rm HI}^{\rm host}$ was used. 

For the diffuse IGM we used  Lorentzian (or the Voigt profile when convolved with the Gaussian velocity distribution), 
and the two-level approximation formula by \citet{Peebles:1993aa} and
\citet{mr98b}, with neutral hydrogen fraction $f_{\rm HI}$ in a redshift range from $z_{\rm IGM,l}=5.67$ and 
$z_{\rm IGM,u}=[z_{\rm GRB},free]$. Also, the GP optical depth is $\tau_{\rm GP}=3.97\times 10^5 f_{\rm HI}[(1+z)/7]^{3.2}$ calculated from the cosmological parameters of \citet{Komatsu:2011aa} and the primordial helium abundances of \cite{Peimbert:2007aa}.
These components can be seen in Figure\,4.
For this exercise we considered the reionization to be uniform (rather then, e.g., patchy). 

Regions of the spectrum cleaned of absorption features like \CII\ or \NV, as well as strong skyline
emission, were excluded by the fit and simple chi-squared minimization
statistics were employed in order to find the best model.
The best-fit model parameters are listed in Table 1 of \citet{Totani:2014aa}, and provide a neutral 
hydrogen fraction of $f_{\rm HI}=0.086^{+0.012}_{-0.011}$.
Furthermore, we were able to exclude with high-confidence a model that contains only a host DLA, while one or  
two components (diffuse IGM only, or IGM+intervening DLA) are equally valuable good fits (Figure\,5).
The presence of a second \Lya\ feature at lower redshift ($z=5.806$) is supported by several metal lines at such redshift, but
further analysis of the possible presence (and strength) of the expected \Lyb\ line allows to exclude that this is another
DLA along the line of sight.

We point out that the chances of finding an intervening DLA at lower $z$ is very low: from the $z\sim5-6$ DLA numbers density the chance of random encountering such system is roughly $\sim3\%$ \citep[][]{Songaila:2010aa,Castro-Tirado:2013aa}. 
Finally, the GRB\,130606A spectrum shows the power of using high-$z$ GRBs to probe the diffuse IGM down to
$f_{\rm HI}\approx0.1$ for gas close ($\leq 5$ Mpc in proper distance) to the host galaxy. This value is consistent with the limits 
obtained by QSO studies \cite[e.g][]{Mortlock:2011aa}.
Such analyses will be even more productive with the upcoming 
generation of 30m telescopes, which will be optimized for  rapid follow-up of such transients as well as be equipped
with powerful NIR spectrographs.

It is interesting that other groups arrived at different results from that
of Subaru presented above. \citet{Chornock:2013aa}  found no evidence for
nonzero IGM \HI, though their upper limit is not seriously in contradiction
with the best-fit IGM \HI\ of \citet{Totani:2014aa}. 
It should be noted that \citet{Chornock:2013aa} used an intrinsic afterglow 
spectral index of $\beta = -1.99$, which is not supported
by observed afterglow colors or the VLT spectrum that shows $\beta = -1.0$.
However, \citet{Hartoog:2015aa} 
found the best fit $f_{\rm HI}$ to be zero, with a stringent $3\sigma$ upper limit of 
$f_{\rm HI}\leq 0.05$, that is statistically inconsistent with the best-fit 
$f_{\rm HI}$ of \citet{Totani:2014aa}.
\citet{Totani:2015aa} examined the origin of this discrepancy
by analyzing the VLT data with the code adopted to the Subaru data.
It is found that the same result as \citet{Totani:2014aa} is obtained
by adopting the same analysis method to the VLT spectrum, and hence
the origin of the discrepancy is likely the difference of analysis methods,
especially the wavelength ranges used. It is important to carefully
optimize an analysis method to minimize systematic uncertainties,
when very high precision spectra are obtained for high-$z$ GRBs.

\section{Escape fraction and re-ionization: first galaxies}
\label{sec:4}
We mentioned in the previous section that a transition from a cold neutral IGM to the hot ionized one must have occurred
at some point in the Universe's history.
A key question is whether this phase change  was predominantly driven
by ionizing radiation from early generations of stars (UV emission from star forming galaxies), or whether other mechanisms 
(e.g. decaying particle ionizing background, contribution from accretion luminosity)
must have had a significant effect.
Considering the known star formation budget at $z\sim6$--9, recent calculations confirm that
it is plausible that massive stars provided sufficient ionizing photons, providing that a substantial fraction
(typical analyses suggest $f_{\rm esc}$ of at least 10--20\%) could escape the host galaxies in which the stars reside
\cite[e.g., ][]{Bouwens:2015aa}.
Unfortunately it is highly challenging observationally  to pin down the escape fraction even at
lower redshifts, and most recent studies have concluded that it is likely to be rather low, e.g. a few
percent or less at $z\sim3$ \citep[e.g., ][]{Grazian:2015aa}.

GRBs provide a novel route to addressing this question, since optical afterglow spectroscopy
of $z>2$ GRBs frequently provides a measurement of the neutral hydrogen column 
density along the line of sight through the host galaxy.
Typically these columns are high, as measured in damped Lyman-$\alpha$ absorbers, and are a lower limit to the column 
that would have been in front of the GRB progenitor, in as much as hydrogen proximate
to the burst may well be ionized by the burst itself.
Thus most lines of sight are opaque to ionizing radiation, and
in particular, very few GRB afterglows have ever been found with columns low
enough to allow even a small fraction of ionizing radiation to escape and therefore to be measured directly
(e.g. via far-UV photometry).

There are reasons to think that at higher redshifts, with more star formation occurring in smaller
galaxies, the escape fraction may increase, although measurements of \nh\ from GRBs at
$z>4$ show at best a modest decline \citep{Thone:2013aa,Chornock:2014aa}. 
Nevertheless, recently, thanks to the discovery of many $z>2$ GRBs
with spectroscopic redshift and \nh\ column measurements, $f_{\rm esc}$ has been measured using an indirect technique, which involves only the determination of \nh\ distribution and it is not subject to systematics due background subtraction \citep{Chen:2007aa,Gnedin:2008aa,Fynbo:2009lr}.

Following the work of \citet{Chen:2007aa}, the mean escape fraction of Lyman limit photons 
averaged over all directions (GRB sightlines) is evaluated according to

\begin{equation}
\langle\fesc\rangle=\frac{1}{n}\sum_{i=1}^{i=n}\exp[-\sigma_{\rm LL}\,N_i({\rm H\,I})],
\end{equation}

where the sum extends over the total number of $n$ GRB sightlines in
the sample and $\sigma_{\rm LL}=6.28\times10^{-18}{\rm cm}^2$ is the photoionization cross section of hydrogen atoms.
They find  $\langle\fesc\rangle=0.02\pm 0.02$ (where error
is estimated using the bootstrap re-sampling method).  They also where able to
determine a 95\% c.l. upper limit of $\langle\fesc\rangle \le 0.075$.

\citet{Fynbo:2009lr}, using a larger number of sight lines selected in a more unbiased way, found very similar results.
These results are also consistent with measurements from Lyman Break Galaxies at $z\sim3$, therefore suggesting that
the \fesc\ from QSOs and sub-$L^*$ galaxies, like the ones hosting GRBs, contribute for a comparable amount 
of ionizing radiation.


The large GRB sample detected by \swift, and accurate redshift measurement from ground based follow-up, 
have sparked deep observations of the highest redshift sample of GRB host
galaxies, in particular near the time when re-ionization was complete. 
Numbers remain small at $z>6$, so the constraints do not yet provide a critical problem for re-ionization,
but samples with good spectroscopy will hopefully improve in the coming years.


\begin{figure*}[t!]
 \includegraphics[width=0.75\textwidth]{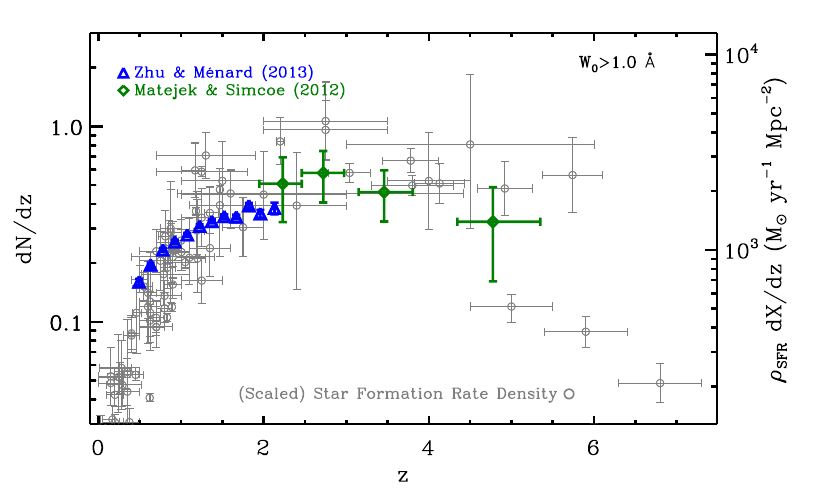}
\caption{Incident rate $dN/dz$ of very strong \MgII\ absorbers ($W_{2796} >1.0$ \AA). Blue points are from 
\citet{Zhu:2013aa} based on SDSS data at moderate redshifts, while the green points show near-infrared measurements from \cite{Matejek:2012bh}. On the right axes we indicate the cosmic star formation rate density, $\dot{\rho}_{*} = dM_{*}/dtdAdX$, scaled by $dX/dz$, where A is comoving area and X is comoving distance.}
\label{fig:1}       
\end{figure*}

\begin{figure*}[t!]
 \includegraphics[width=0.75\textwidth]{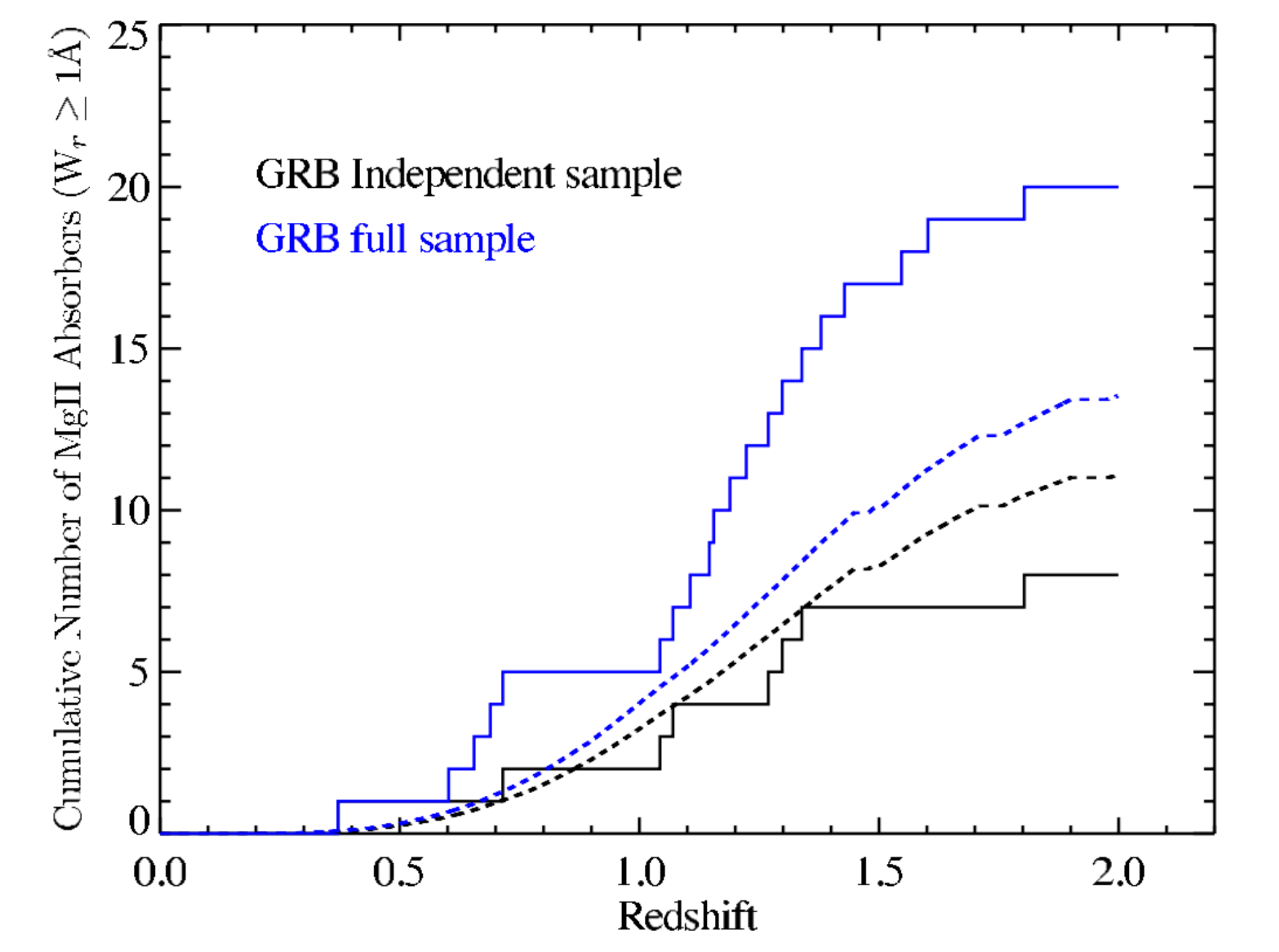}
\caption{Cumulative distribution of strong \MgII\ absorbers along GRB sight
lines for an independent sample from \citet{Prochter:2006fr} (Sample I) and for the full sample
(Sample F, black and blue solid curves, respectively).
These are compared to the predicted incidence based on measurement along
QSO lines of sight (dashed curves). The independent Sample I actually shows
fewer absorbers than expected while a modest excess remains in Sample F.
Neither result corresponds to a statistically significant difference from the QSO
results.}
\label{fig:2}       
\end{figure*}

\begin{figure*}[t!]
 \includegraphics[width=0.75\textwidth]{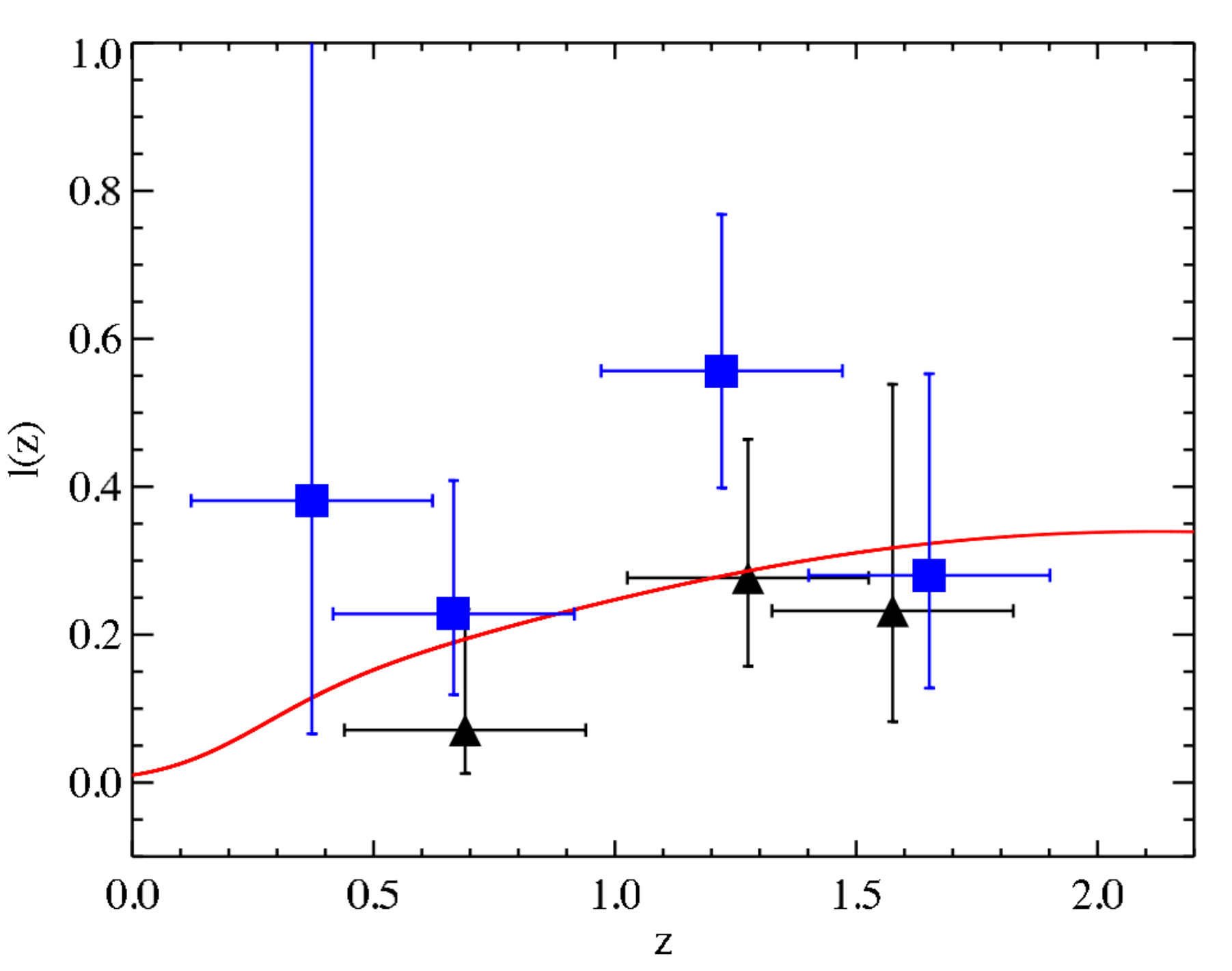}
\caption{Incidence of very strong \MgII\ absorption intervening systems, $dN/dz$ or \loz, evolution  for the
sample of GRB sight lines: triangles and square symbols refer to the Sample I
and Sample F, respectively \citep[see ][ for more details]{Cucchiara:2013aa}. The red curve shows the evolution of the 
\MgII\ incidence along quasar sight lines as recently computed by \cite{Zhu:2013aa} . 
We derive an average \loz = 0.20 for Sample I, in agreement with the
prediction, while \loz= 0.36 for Sample F, indicating a slight overabundance
of absorbers compared to the QSOs.}
\label{fig:3}       
\end{figure*}

\begin{figure*}[t!]
 \includegraphics[width=0.75\textwidth]{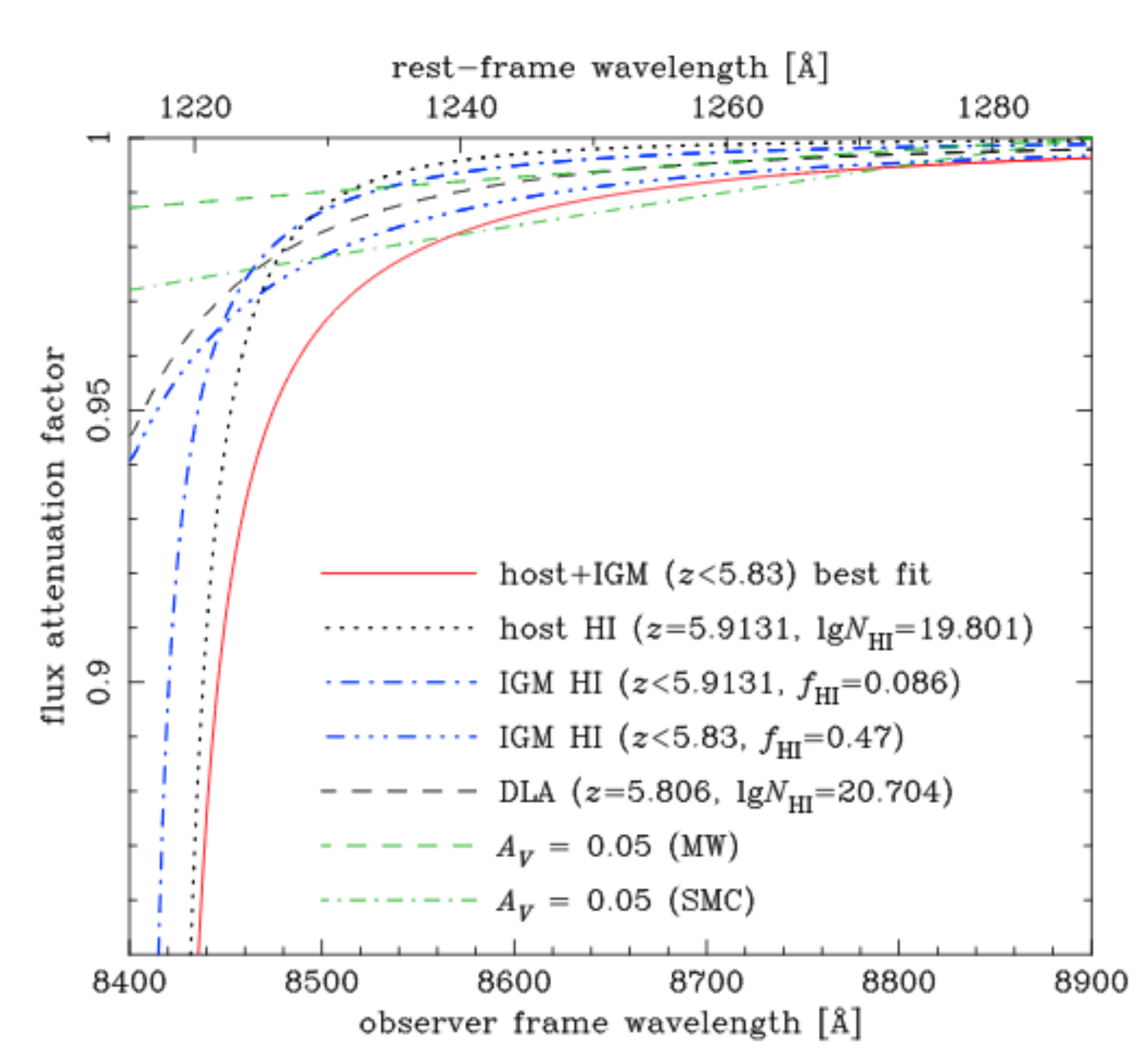}
\caption{Flux attenuation factor of the \Lya\ damping wing by various \HI\ components and extinction at the host. The red solid curve is the total
 attenuation by \HI\  in IGM and the host galaxy. The IGM and host components of this model are also separately shown. The IGM absorption model and the DLA component of the intervening DLA model are also shown for comparison. The green curves are the effect of introducing extinction by dust in the host galaxy (normalized to unity at 8900 \AA), using the MW or SMC extinction curves \citep{Fitzpatrick:1999aa,Gordon:2003aa}. }
\label{fig:4}       
\end{figure*}

\begin{figure*}[t!]
 \includegraphics[width=0.75\textwidth]{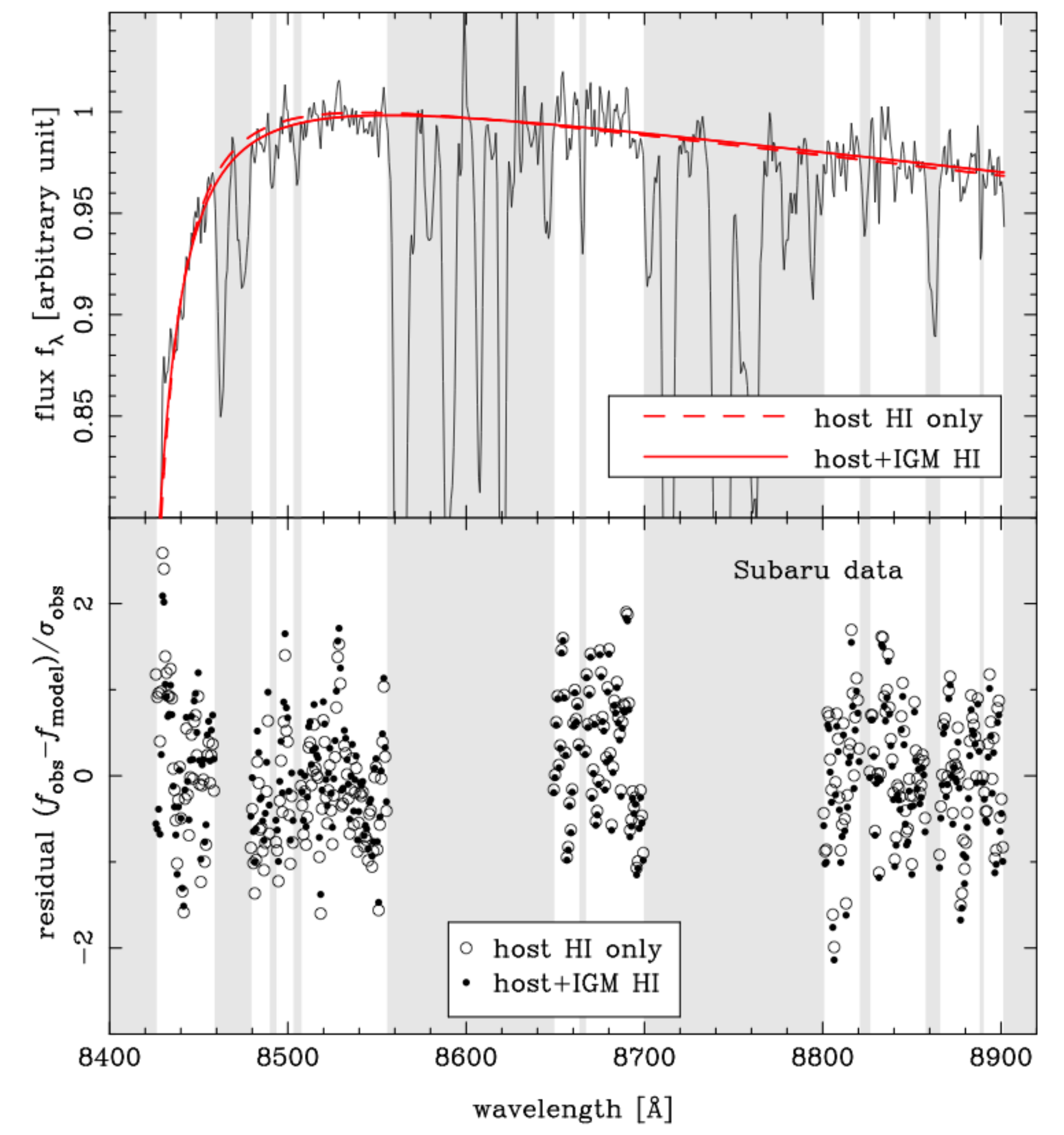}
\caption{\emph{Top:} The optical spectrum of GRB\,130606A afterglow taken by the SUBARU telescope, around the redshifted \Lya\ and its red damping wing region. The best-fit curves of the host \HI\ only model and the host+IGM model are shown by thick dashed and solid curves, respectively.
\emph{Bottom:} The fit residuals ($f_{\rm obs} - f_{\rm model} )/\sigma_{\rm obs}$ are shown for the two models. The gray shaded regions indicate the wavelength ranges removed from the fits because of apparent features of absorption lines or airglow.}
\label{fig:5}       
\end{figure*}



\bibliographystyle{aps-nameyear}     
\bibliography{journals_apj,bibthesis,MgII,bibmaster,bibi090429B,bib_master}   

\end{document}